# From rings to bulges: evidence for rapid secular galaxy evolution at z~2 from integral field spectroscopy in the SINS survey[1]

Genzel, R.[1,2], Burkert, A.[3], Bouché, N.[1], Cresci, G.[1], Förster Schreiber, N.M.[1], Shapley, A.[4], Shapiro, K.[5], Tacconi, L.J.[1], Buschkamp, P.[1], Cimatti[6], A., Daddi[7], E., Davies, R.[1], Eisenhauer, F.[1], Erb, D.K.[8], Genel, S.[1], Gerhard, O.[1], Hicks, E.[1], Lutz, D.[1], Naab, T.[3], Ott, T.[1], Rabien, S.[1], Renzini, A.[9], Steidel, C.C.[10], Sternberg, A.[11] & Lilly, S.J.[12]

[1] *Max-Planck-Institut für extraterrestrische Physik (MPE), Giessenbachstr.1, D-85748 Garching, Germany (genzel@mpe.mpg.de)*

[2] *Department of Physics, Le Conte Hall, University of California, Berkeley, CA 94720, USA*

[3] *Universitäts-Sternwarte Ludwig-Maximilians Universität (USM), Scheinerstr. 1, München, D-81679, Germany*

[4] *Department of Physics and Astronomy, 430 Portola Plaza, University of California, Los Angeles, CA 90095-1547, USA*

[5] *Department of Astronomy, Campbell Hall, University of California, Berkeley, CA 94720, USA*

[6] *Istituto Nazionale di Astrofisica – Osservatorio Astronomico di Bologna, Via Gobetti 101, I-40129 Bologna, Italy*

[7] *Service d'Astrophysique, Dapnia CEA, Saclay, France*

[8] *Harvard-Smithsonian Center for Astrophysics, 60 Garden Street, Cambridge, Mass. 02138, USA*

[9] *Osservatorio Astronomico di Padova, Vicolo dell'Osservatorio 5, Padova, I-35122, Italy*

[10] *California Institute of Technology, MS 105-24, Pasadena, CA 91125, USA*

[11] *School of Physics and Astronomy, Tel Aviv University, Tel Aviv 69978, Israel*

[12] *Institute of Astronomy, Department of Physics, Eidgenössische Technische Hochschule, ETH Zürich, CH-8093, Switzerland*

---

[1] Based on observations at the Very Large Telescope (VLT) of the European Southern Observatory (ESO), Paranal, Chile

# Abstract

We present Hα integral field spectroscopy of well resolved, UV/optically selected z~2 star-forming galaxies as part of the SINS survey with SINFONI on the ESO VLT. Our laser guide star adaptive optics and good seeing data show the presence of turbulent rotating star forming rings/disks, plus central bulge/inner disk components, whose mass fractions relative to total dynamical mass appears to scale with [NII]/Hα flux ratio and 'star formation' age. We propose that the buildup of the central disks and bulges of massive galaxies at z~2 can be driven by the early secular evolution of gas-rich 'proto'-disks. High redshift disks exhibit large random motions. This turbulence may in part be stirred up by the release of gravitational energy in the rapid 'cold' accretion flows along the filaments of the cosmic web. As a result dynamical friction and viscous processes proceed on a time scale of <1 Gyr, at least an order of magnitude faster than in z~0 disk galaxies. Early secular evolution thus drives gas and stars into the central regions and can build up exponential disks and massive bulges, even without major mergers. Secular evolution along with increased efficiency of star formation at high surface densities may also help to account for the short time scales of the stellar buildup observed in massive galaxies at z~2.

## *1. Introduction*

Deep surveys have become efficient in detecting z~1.5-3.5 star-forming galaxy populations near the peak of the cosmic star formation and QSO activity (e.g. Steidel et al. 1996, 2004, Franx et al. 2003, Daddi et al. 2004b). Large samples are now available, based on their rest-frame UV magnitude/color properties (the so-called 'BX/BM' criterion: Adelberger et al. 2004, Steidel et al. 2004), or on rest-frame optical magnitude/color properties ('star-forming', or 's'-BzKs: Daddi et al. 2004a,b, DRGs: Franx et al. 2003). These selection criteria sample fairly luminous (L~$10^{11-12.5}$ $L_\odot$) galaxies with star formation rates of 10-300 $M_\odot$/yr, with a range of ages (10 Myrs - 3 Gyrs) and stellar masses ($M_*$~$10^{9-11.5}$ $M_\odot$) (Shapley et al. 2005, Förster Schreiber et al. 2004, Erb et al. 2006b,c, Daddi et al. 2004a,b). These galaxies contribute a large fraction of the cosmic star formation activity and stellar mass density at z~2 (Reddy et al. 2005, Rudnick et al. 2005, Grazian et al. 2007).

Open issues are how these high-z galaxies formed their stars, when and how the first disks and bulges formed and how their subsequent evolution was related to each other. Was the star formation history, and especially the formation of bulges driven by short, intense bursts following major mergers (mass ratios <3:1)? Or was this evolution dominated by rapid but more continuous 'cold' accretion of gas from the dark matter halos, including large mass ratio (≥10:1) gas rich minor mergers, as proposed by Semelin & Combes (2002), Birnboim & Dekel (2003) and Kereš et al. (2005)? Motivated by the cold dark matter paradigm of structure formation, the (major) merger interpretation has been favored by many for some time, especially for bulge formation. Recent observations and simulations have lent support for the 'rapid accretion', or 'cold flow' scenario, at least for the formation of high-z disks.

Optically/UV-selected star-forming galaxies at z~0.5-2.5 exhibit a fairly tight relationship between stellar mass and star formation rate, arguing against a significant contribution from and time spent in short luminous bursts, but rather favoring a high duty cycle of steady star formation, at least for these populations (Noeske et al. 2007, Elbaz et al. 2007, Daddi et al. 2007). Imaging studies have revealed the presence of a number high-z disk galaxies (e.g. Labbé et al. 2003, Stockton et al. 2008). The first spatially resolved studies of the ionized gas kinematics in optically/UV-selected, z~2 star-forming galaxies have yielded a surprisingly high abundance of large rotating disks among the more massive ($M_{dyn}$~$10^{11}$ $M_\odot$) and restframe optical bright ($K_s \leq 20$) systems. In this sub-sample only a smaller fraction (~1/3) are obvious major mergers (Shapiro et al. 2008, Förster Schreiber et al. 2006, Genzel et al. 2006, Wright et al, 2007, Law et al. 2007, Cresci et al. 2008). In contrast to these UV/optically selected galaxies, the very luminous submillimeter galaxies (SMGs) at z~1-3 appear to mostly be major mergers in various stages of evolution (Tacconi et al. 2006, 2008, Swinbank et al. 2006). The most recent generation of cosmological galaxy evolution simulations, based on dark matter merger tree simulations, find that smooth accretion of gas and/or minor mergers dominate high-z star formation, while the overall mass assembly of galaxies, and especially the most massive ones, is dominated by mergers (Bower et al. 2006, Kitzbichler & White 2007, Naab et al. 2007, Guo & White 2008, Davé 2008, Genel et al. 2008). This conclusion seems to hold for simulations of the baryonic physics based on semi-analytic recipes, as well as for hydrodynamic simulations. Taken together, these recent observational and theoretical results are beginning to form a well formulated framework for the formation of disks that needs now to be tested by detailed 'in situ' observations. Such measurements also need to explore the evolution of the first disks, as well as the formation bulges.

In this paper we present and analyze high-quality SINFONI/VLT integral field (IFU) spectroscopy (Eisenhauer et al. 2003, Bonnet et al. 2004) of 8 representative UV/optically selected (BX/s-BzK) z~2 star forming galaxies. The data for two of these galaxies are discussed here for the first time. For these galaxies we employed laser guide star (LGSF) adaptive optics (AO) to improve the angular resolution. The data for the other 6 galaxies are discussed in other publications of our team (Förster Schreiber et al. 2006, 2008a, Genzel et al. 2006, Cresci et al. 2008) and we present here a new, detailed analysis. We show that these data deliver interesting new constraints on the issues of disk evolution, bulge formation and star formation processes. We adopt a $\Lambda$CDM cosmology with $\Omega_m$=0.27, $\Omega_b$=0.046 and $H_0$=70 km/s/Mpc and a Chabrier (2003) initial stellar mass function.

## *2.    Observations and Analysis*

### 2.1 Data

As part of the SINS survey of high-z galaxy kinematics we observed the H$\alpha$ and [OIII] lines and the rest-frame R-band continuum in the UV selected BX galaxy Q2346-BX482 (z=2.26, Erb et al. 2006b,c, Förster Schreiber et al. 2006), as well as the H$\alpha$ line and restframe R-band continuum in the optically selected s-BzK galaxy BzK6004-3482 (henceforth BzK6004, z=2.39, Kong et al. 2006). For a more detailed description of the SINS/SINFONI sample we refer to Förster Schreiber et al. (2008a). Briefly, the SINS-BzK targets are drawn directly from the list of optically selected K$\leq$20 s-BzK galaxies (in the case of the Deep 3a field: Kong et al. 2006). The SINS-BX/BM sample is a fair representation of the larger Erb et al. (2006a) H$\alpha$ sample in terms of source sizes and dynamical masses. However, our additional selection criteria emphasize somewhat brighter (<F(Hα)> $10^{-16}$ compared to $6\times10^{-17}$ erg/s/cm$^2$) and somewhat broader line width (<$v_c$>~175, ±68 km/s sample dispersion) systems than the average galaxy in the larger sample of Erb et al. (2006b,c, <$v_c$>~140). For BX482 and BzK6004 we used SINFONI (Eisenhauer et al. 2003, Bonnet et al. 2004) on UT4 at the ESO VLT in the seeing-limited and LGSF AO modes (Rabien et al. 2003, Bonaccini Calia et al. 2006). In addition we re-analyzed the seeing-limited K-data in the three BX galaxies Q2343-BX389 (z=2.174), Q2343-BX610 (z=2.211) and SSA22-MD41 (z=2.172) (Förster Schreiber et al. 2006, Cresci et al. 2008). Table 1 gives the observing log and lists the FWHM angular resolutions of these data. The FWHM spectral resolution of all K-data cubes is about 80 km/s. For part of our analysis we also refer to results of three additional SINS z~2 BX/s-BzK galaxies for which we have LGSF and natural guide star (NGS) AO SINFONI observations with resolutions of 0.2". These galaxies are Q1623-BX502 (z=2.16, Förster Schreiber et al. 2006), BzK15504 (z=2.38, Genzel et al. 2006) and ZC782941 (z=2.18, SINS and COSMOS teams in prep.). For a description of our general data reduction methods we refer to Schreiber et al. (2004) and Förster Schreiber et al. (2006). With the final data cubes in hand, we smoothed the data by two spatial pixels and then fitted Gaussian line profiles to each pixel. To blank out non-meaningful or low signal-to-noise regions, we multiplied the resulting velocity and velocity dispersion maps with a mask constructed from the total Hα line intensity distribution.

The five main sources we discuss in this paper were selected from the SINS survey of z~1.5-3 star forming galaxies (Förster Schreiber et al. 2008a) on two criteria. First they are among those galaxies for which quantitative kinemetry of their velocity fields shows that they are very likely rotating disks and are not undergoing major mergers (Shapiro et al. 2008). Second among the SINS rotating disks these five galaxies are those with the highest quality data and are the best spatially resolved systems (FWHM major axis diameter ~ 3 to 7.5 times the FWHM instrumental resolution). This makes possible a quantitative and fairly robust analysis of the dynamics, including for the first time the determination of the ratio of the dynamical mass in the central few kpc to the total dynamical mass. As a result the galaxies we have chosen belong to the largest (30 percentile) and most massive (50 percentile) members of the SINS sample but otherwise are representative of other massive (~$10^{11}$ $M_{\odot}$) star-forming galaxies at z~2.

Of the three additional SINS galaxies discussed in this paper, BzK15504 and ZC782941 are similar to the other five large BX/s-BzK galaxies discussed above, in terms of size, mass and dynamical properties. In contrast BX502 is a representative of the fairly common class of compact, young and somewhat lower mass members among the BX population in the Erb et al. (2006b,c) sample, a number of which appear to be 'dispersion dominated', where the random velocity dispersion is comparable to any large scale rotational component (Law et al. 2007). For these three additional galaxies we also resolve the disk dynamics but the data are unsuitable for a bulge analysis. The central velocity field of BzK15504 shows an outflow potentially triggered by an AGN, The outer velocity field of ZC782941 is disturbed by an ongoing minor merger. BX502 is too compact for a detailed dynamical analysis.

**2.2 Modeling**

We fit the Hα velocity fields of all galaxies with simple rotating disk models (Förster Schreiber et al. 2006, Cresci et al. 2008). These disk models compute data cubes from input structural parameters. A specific model requires specifying one or several mass/light components, parameterized by azimuthally symmetric analytic functions (e.g. exponential disks, Gaussians, rings etc.), the total dynamical mass $M_{dyn}$, z-scale height, inclination, position angle of major axis on the sky and a component of constant velocity dispersion throughout the disk ($\sigma_{01}$). The model data are then convolved with the angular and spectral resolution profiles and sampled at the observed pixel scales. The total dynamical mass and mass distributions then fully define the velocity field, which is computed assuming spherical symmetry. Given the large z-scale heights $h_z$ inferred for the z~2 galaxies (see below) this approximation holds at the 10-15% level of velocity estimation (Binney & Tremaine 2008, chapter 2). This is sufficient for our modeling. A given z-scale height results in an additional z-velocity dispersion $\sigma_{02}$. In the approximation of a very large and very thin disk this is given by

$$\frac{h_z}{R} = 0.5 \left(\frac{\sigma_{02}}{v_d}\right)^2 \qquad (1a),$$

where $v_d(R)$ is the rotation velocity at radius R (Binney & Tremaine 2008, eq. 4.302c). For a thick or a compact disk equation (1) needs to be replaced by

$$\frac{h_z}{R} = \frac{\sigma_{02}}{v_d} \qquad (1b).$$

The total z-velocity dispersion in our models then is $\sigma_0 = \sqrt{\sigma_{01}^2 + \sigma_{02}^2}$.

To determine the model parameters that best describe each galaxy, we constructed a first order input model with parameters for the light/mass distribution and inclination as estimated from PSF-corrected Gaussian major and minor axis FWHM sizes of the integrated Hα emission. We then refined these models iteratively to match the velocity and velocity dispersion maps for each galaxy. We determined the best fitting galaxy parameters from $\chi^2$-minimization between our disk model outputs and the two-dimensional velocity and velocity dispersion maps. In the case of BX482 we included in this $\chi^2$-miminization the extinction corrected, two-dimensional Hα light distribution in the optimization (see below). For the other galaxies the Hα brightness distribution is too asymmetric and clumpy as to be useable as a detailed constraint beyond that obtained from the initial major axis size and aspect ratio above. While the disk orientation, inclination and kinematic parameters (rotation curve, velocity dispersion distribution) can be well determined from the data sets, the details of the structural parameters (disk scale length, ring widths, size of the central mass concentration etc.) are necessarily less well and not uniquely constrained.

The key new parameter we are able to extract from the high quality data discussed in this paper (as compared to the work of Förster Schreiber et al. 2006 and Cresci et al. 2008) is a ***'mass concentration parameter'***, that is the ratio of the dynamical mass in the center (we chose R ≤ 0.4"~3.3 kpc) to that in the overall observed disk to R ≤ 1.2" (~10 kpc). This ratio can be interpreted as the mass in a central bulge or central disk to the overall disk mass. To derive this mass concentration parameter we repeated the velocity field $\chi^2$-miminization along a major axis cut with the $M_{dyn}(\leq 0.4")/M_{dyn}(1.2")$ ratio as the only free variable, once the best fitting values of inclination, positional angle, total mass and structural parameters (ring radius and width, or disk scale length) were determined from the 2d $\chi^2$-miminization. Along this axis the rotation velocity and velocity dispersion are maximally sensitive to the rotation curve, and thus to the mass distribution (van der Kruit & Allen 1978). This is especially true for the central regions, which are poorly resolved by our data. Our observations only provide a modest number of independent 2d resolution elements across the observed galaxies. In this regime the 2d velocity data off the major axis mainly serve to constrain the overall orientation of the galaxy (inclination and major axis position angle). The regions near the minor axis also are sensitive to non-circular motions since the rotation component is minimized there (Figure 1 of van der Kruit & Allen 1978). For the determination of the mass concentration parameter $M_{dyn}(\leq0.4")/M_{dyn}(1.2")$ inclination and position angle drop out and thus to first order do not enter the error budget. As a result we can fit for this single parameter from the major axis data once the overall disk geometry is known.

For each galaxy we considered two types of models. One is an exponential disk model of constant dynamical mass to Hα light ratio and radial scale length $R_d$. The other is a combination of a Gaussian ring (at radius $R_r$ and FWHM width $2\Delta R$), plus a central mass (again a Gaussian at R=0 and $\Delta R$=0.15-0.2"). Such a ring can also generically describe a flat surface brightness disk when $R_r \sim \Delta R$.

With some exceptions described below the velocity and velocity dispersion fields in the five galaxies discussed here are remarkably well matched by simple disk models. The maximum residuals from the best-fit velocity fields range between 10 and 50 km/s~0.05-0.2 $v_d$(max). The most serious limitation of our modeling comes from the clumpiness and asymmetry of the Hα brightness distributions in most of the program galaxies, which limits the detail that can be derived on the matter distribution.

## *3. Results*

Figures 1 through 8 summarize the data, models and data-model residuals for the five main SINS z~2 star-forming galaxies we are discussing in this paper. We also show comparisons between the integrated Hα line maps and the broad-band stellar continuum maps, extracted either from the SINFONI cubes, or from the HST NICMOS/NIC2 imaging through the F160W filter (hereafter: "$H_{160}$', Förster Schreiber et al. 2008b). Tables 2 and 3 summarize the derived physical properties of these five galaxies, and in particular the mass concentration parameter $M_{dyn}(\leq0.4")/M_{dyn}(1.2")$ introduced in the last section. In addition to the basic structural and dynamical parameters deduced from our modeling, we also list extinction corrected star formation rates inferred from Hα (based on the Calzetti 2001 recipes), stellar masses inferred from modeling of the rest-frame UV/optical spectral energy distribution (see Förster Schreiber et al. 2008a for details) and gas masses estimated from the star formation rates and the universal Kennicutt-Schmidt star formation. For the latter we adopt the gas mass-star formation relation proposed by Bouché et al. (2007). Finally we also list in Table 3 the [NII]/Hα flux ratio, rest-frame equivalent width of Hα and star formation surface densities.

In the next sections we first discuss the individual galaxies, starting with the new laser guide star AO observations. None of these five galaxies show evidence for an ongoing or recent major merger. Their star formation rates nevertheless are quite high. From Table 2 one sees that the time required to exhaust the present gas in these galaxies at the present star formation rate is ~2 x$10^8$ years, an order of magnitude smaller than the interval of cosmic times (~2 Gyr) over which these galaxies are selected (1.4<z<2.5), or the stellar ages determined from the SED modeling (see however the caveats in 4.7). Because of the quoted tight correlation of star formation rate with stellar mass (Daddi et al. 2007), the high star formation rates typical of these z~2 galaxies must clearly be sustained for most of this time interval. Given that the consumption time is a factor ~10 less, this suggests that star formation in our study galaxies is sustained by continuous inflow of cold gas in a quasi-steady-state regime, in good agreement with the conclusions of Erb (2008).

### 3.1 Q2343-BX482

In our initial seeing-limited SINFONI Hα data (Förster Schreiber et al. 2006) this galaxy appeared to be 'tadpole'-shaped with a bright emission spot to the south-east and a conical open emission region to the north-west. Our higher resolution LGSF data as well as the NIC2 $H_{160}$ image of Förster Schreiber et al. (2008b) shown in Figure 1 now reveal that the Hα and rest frame V-band continuum emission forms a complete, clumpy ring of radius ~0.8", with a strong east-west asymmetry in the brightness distribution, but with a coherent kinematic structure. The additional seeing limited data of the 5007 Å [OIII] line exhibit an even stronger intensity asymmetry but the same kinematics. As a result the [OIII]/Hα flux ratio from the seeing limited mode for comparable resolution exhibits a smooth east-north-east to west-south-west gradient (Figure 1 top right panel). There is either a large scale gradient in excitation/abundance or in extinction. An abundance gradient is not very likely since the [NII]/Hα ratio is constant across the source, to within the measurement uncertainties. If there is an east-west extinction gradient, the intrinsic Hα distribution has to be more symmetric. The extinction gradient could be caused by diffuse foreground dust, or by dust partially mixed in with the Hα emission in the ring. We have used the observed [OIII]/Hα gradient to correct for the differential extinction across the source. Adopting an LMC/SMC extinction curve and a 'foreground screen' extinction model[2], the south-east to north-west asymmetry in the Hα brightness distribution disappears and in fact is reversed (top left panel of the major axis cut in Figure 3). The extinction corrected two-dimensional brightness distribution is remarkably well matched by a symmetric inclined ring (inclination $65\pm7^0$, position angle $-65\pm5^0$ (positive= east of north)). The maximum differential extinction across the entire galaxy is $\Delta A_V \sim 1.9$. The average residual (data-model) brightness across the galaxy is 9 % of the maximum surface brightness, with a few regions deviating by up to 40%.

The velocity and velocity dispersion data in Figures 2 & 3 are well fit either by a superposition of a fairly narrow ($2\Delta R \sim 0.3"$) ring at radius $R_r=0.85"\pm0.1"$, plus a central mass (a bulge or central disk), or by an exponential disk model (upper right inset in Figure 3). The former model is obviously motivated by the morphology of the Hα emission, while latter is the canonical distribution found in low-z disk galaxies. For both models the maximum intrinsic rotation velocity is $v_d=235\pm40$ km/s (middle right panel of Fig.3). Average residuals in the velocity and velocity dispersion maps over most of the galaxy are <10 km/s, with larger deviations in velocity (up to 100 km/s) along the northern edge of the galaxy (top right panel of Figure 2). To match the steep slope of the major axis rotation curve and also the peak of velocity dispersion near the kinematic center (bottom left panel of Figure 3), the ring model requires an additional central mass, either a bulge and/or a concentrated inner disk (HWHM~0.3"). The best fitting mass of the central component within 0.4" is $2.9\pm0.3 \times 10^{10}$ M$_\odot$, implying a mass concentration parameter $M_{dyn}(\leq 0.4")/M_{dyn}(1.2")=0.205\pm0.03$. With this ring (+bulge) input model the major axis Hα intensity, velocity and velocity dispersion data are fit with $\chi_r^2$~1. A ring model without a central mass component can be excluded with high significance. An exponential disk model does as well as the ring (+bulge) model in fitting the dynamical data and implies approximately the same mass concentration parameter as the ring (+bulge) model (~0.16, bottom right inset of Figure 3). However, the exponential disk model obviously fails to match the ring morphology of the Hα (and H-band continuum) maps in Figure 1. If the intensity of Hα traces the mass distribution a constant mass to light ratio, an exponential disk model is excluded with high significance in the fitting.

---

[2] in a screen extinction model the line intensity at wavelength λ depends on dust opacity $\tau_\lambda$ at λ as $\exp(-\tau_\lambda)$. For an LMC/SMC extinction curve appropriate for the sub-solar abundance implied by the BX482 [NII]/Hα line ratio $\delta=\tau_{[OIII]}/\tau_{H\alpha} \sim 1.5$ (Calzetti 2001). The differential extinction correction to the Hα data then scales as $\{F([OIII])/F(H\alpha)\}^{-\beta}$, with $\beta=(\delta-1)^{-1} \sim 2$.

The obvious question is whether the ring morphology of the Hα emission may just trace the regions of currently active star formation and not the overall mass, and may thus just represent 'frosting on the cake'. The middle left panel of Figure 3 shows that the projected rotation velocity rises locally upward at $R\leq R_r$ and then appears to fall at $R>R_r$ on the eastward side. This is consistent with our assumption that the ring traces the dynamical mass. More importantly the NIC2 image (rest frame ~V-band) shows that the visible stellar distribution is also dominated by the ring structure (middle inset of Figure 1, Förster Schreiber et al, 2008b). The integrated $H_{AB}$-band magnitude of BX482 is 22.4 (Förster Schreiber et al. 2008b). Adopting a constant star formation model with Z=0.4 Z$_\odot$ tracks, a Kroupa/Chabrier IMF and a Calzetti (2001) extinction model with $A_V$~1.5-2, this magnitude implies a mass of live stars of 4-9x10$^{10}$ M$_\odot$ for a plausible age range of 0.3-1 Gyrs. Adding this stellar mass and the gas mass estimated from the Hα-based star formation rate and a Kennicutt-Schmidt recipe ($3 \times 10^{10}$ $M_\odot$, Table 2) results in a baryonic mass of $7-12 \times 10^{10}$ $M_\odot$ associated with the visible ring (+bulge) structure. Allowing for a 20-40% dark matter contribution to the dynamical mass within ~10-15 kpc (section 4 below) then yields a sum of dark matter and baryonic matter of $10-18 \times 10^{10}$ $M_\odot$. Our dynamical modeling yields a total dynamical mass of $14.3 \pm 2 \times 10^{10}$ $M_\odot$, leaving little or no space for any additional underlying older stellar distribution that is not traced by our data. We conclude that the baryonic matter distribution of BX482 is most likely a ring, plus a small bulge/central disk.

There are significant deviations of the data from the best fitting model. All rotating ring/disk models exhibit significant residuals in the northern part of the ring (upper right panel in Figure 2), perhaps indicating that there is an addition large scale, radial velocity component near the minor axis of the ring with $v_{radial}$ ~85 km/s~0.4 $v_d$. This is reminiscent of the findings in BzK15504 (Genzel et al. 2006). In that case the radial motions may be caused by streaming along an m=2 (bar-like) asymmetry. Further, for a geometrically thin disk the residuals in data-model velocity dispersion are large throughout the ring. An additional random velocity component ($\sigma_0$~55±5 km/s) needs to be introduced everywhere in the ring. As a result a strong conclusion of the modeling is that $v_d/\sigma_0$= 4.3±0.8 (middle panel of right column in Figure 1). This conclusion is equivalent with the requirement in equation (1b) that the gas has a z-scale height of ~1.6 kpc. The disk/ring of BX482 clearly is thick and turbulent.

Table 2 summarizes the derived properties of BX482. Unfortunately there is no multi-band photometry of this system and we cannot determine the stellar age of this system. Based on the approximate estimate of the stellar mass given above, the star formation age, $M_*/SFR$, is relatively small and ranges between 0.3 and 0.6 Gyrs. BX482 also has a low [NII]/Hα ratio for its dynamical mass, which in turn may imply an early evolutionary state and fairly young age (section 4.1, top right inset of Fig.10). In morphology BX482 resembles the young and blue, 'clump-cluster', 'tadpole' or 'chain' galaxies in the HUDF studied by Elmegreen & Elmegreen (2005, 2006) and Elmegreen et al. (2007). Elmegreen et al. (2007) emphasize that a fairly large fraction of their clumpy HUDF galaxies have radial brightness distributions flatter than that of exponential disks (Sersic indices <1), again similar to that of BX482 (Figure 1, Förster Schreiber et al. 2008b). Their brightness distributions do not exhibit a maximum at the geometric centroid. MD41 and BX389 discussed below also are examples of such chain/tadpole galaxies with flat brightness distributions.

### 3.2 BzK6004-3482

The Hα emission in this fairly red and K-bright ($K_s$=18.9) s-BzK galaxy is also well fitted by a ring or disk around a fairly massive central bulge (Figures 4 & 5). That bulge is clearly detected in the SINFONI K-continuum (Figure 4 left column). Proceeding then as for BX482 in the last section, the best fitting model ($\chi_r^2$=1.1) is a rotating wide ring centered at radius $R_r$=0.85" (6.9 kpc) at inclination 35±10$^0$, major axis position angle -18$^0$±10$^0$, and maximum rotation velocity of 255 km/s. Its large inferred width (ΔR~7 kpc) implies that the ring effectively is identical with a flat disk (upper right panel of Figure 5). The central bulge has a mass of 5.0±0.5x10$^{10}$ M$_\odot$, corresponding to $M_{dyn}(\leq 0.4")/M_{dyn}(1.2")=0.37\pm0.04$. The inferred $v_d/\sigma_0$ value (4.3±1.5) is quite sensitive to the adopted inclination. An exponential disk model (dotted lines in Figure 5) is as good as the ring/disk model in terms of the Hα surface brightness distribution but does not match the large central dispersion peak. An exponential disk model also overpredicts the rotation velocity (and thus mass) at R>1" (bottom and middle left panels in Figure 5). A combination of an exponential disk and a bulge fits the data as well as the broad ring (+bulge) model.

The derived parameters of BzK6004 are qualitatively very similar to but more extreme than those of BX482. The mass concentration parameter $M_{dyn}(\leq 0.4")/M_{dyn}(1.2")$ of BzK6004 is almost twice as large as that of BX482. The high K-flux, the red color, the ~2.5 Gyr stellar age derived from population synthesis modeling of its spectral energy distribution (Förster Schreiber et al. 2008a) and the large [NII]/Hα ratio all suggest that BzK6004 is already quite evolved, perhaps also signaling the presence of a central AGN. In many ways it resembles a modern Sa galaxy. Is it possible that the larger mass concentration parameter of BzK6004 is a result of evolution?

### 3.3 Q2343-BX389, Q2343-BX610 & SSA22-MD41

For comparison to these two galaxies studied in LGSF mode we re-analyzed three additional large BX galaxies (Q2343-BX389, Q2343-BX610 & SSA22-MD41) for which our seeing-limited data (Förster Schreiber et al. 2006) are of sufficiently high resolution to be able to place a significant constraint on the mass concentration parameter. Figures 6, 7 and 8 give the major axis cuts and best fitting models for the three BX galaxies. These results are remarkably similar to and consistent with those in the two new sources presented above, although the poorer resolution of these data sets leads to increased uncertainties of the inferred parameters, as listed in Tables 2 and 3.

In SSA22-MD41 the data exhibit no significant evidence for a central mass in this high inclination 'chain galaxy'. The morphology of the NIC2 $H_{160}$ and ACS images (Förster Schreiber et al. 2008b) also suggests a flat extended distribution, consistent with our ring interpretation. The fairly linear rotation curve along the entire major axis of the galaxy, the lack of a central intensity maximum and of a central velocity dispersion peak set an upper limit of 15% (3σ) to $M_{dyn}(\leq 0.4")/M_{dyn}(1.2")$. Similar to the situation in BX482, an exponential disk model is strongly disfavored if the Hα line emission and NIC2 H-band continuum maps trace the overall mass distribution. A constant star formation modeling with parameters similar to those adopted above for BX482 together with the observed H- and K-band AB magnitudes of MD41 ($H_{AB}$=22.8, $K_{AB}$=22.3, Förster Schreiber et al, 2008b) yields a live stellar mass of 1.7-5.6x10$^{10}$ $M_\odot$. Adding this stellar mass to the gas mass estimate of Table 3 (2x10$^{10}$ $M_\odot$) is again consistent with the dynamical mass (7x10$^{10}$ $M_\odot$) once allowance is made for a 20-40% dark matter contribution (section 4). As in the case of BX482 there is little or no space for any additional unseen older stellar component. An exponential disk model thus is not favored for MD41. Likewise the stellar mass estimate above and the extinction corrected Hα-based star formation rate (Table 3) suggest a young star formation age, 0.2-0.5 Gyrs.

BX610 and BX389, on the other hand, have 1-3 Gyr stellar/star formation ages as derived from multi-band fitting to their spectral energy distributions (Förster

Schreiber et al. 2008a). They also exhibit evidence for a central mass concentration constituting about 40% of the dynamical mass in the central 10 kpc. The NIC2 $H_{160}$ image of BX610, in particular, shows a set of bright clumps concentrated within ~0.4" near the dynamical center of the galaxy and surrounded by a lower surface brightness disk of total radius ~ 0.7" (Förster Schreiber et al. 2008b). For BX389 an exponential disk (+bulge) model provides equally good fits to the R<1" kinematics as a wide ring (+bulge) model but probably overestimates the mass at R>1"(Figure 7). For BX610 an exponential disk model formally is even a somewhat better fit to the data than a ring (+bulge) model (Figure 8).

## *4. Discussion*

We have shown in the last section that in the best SINS data sets we can now begin to place constraints on the dynamical mass of a central bulge/inner disk, in addition to investigating in detail the dynamical properties of the surrounding outer proto-disk. In the following section we discuss the results and implications for the sources we have analyzed in 3.1-3.3. We now also include BX502, BzK15504 and ZC782941 for the specific discussion of their random motions.

Essentially the entire dynamical mass in the central 0.4" of all sources must be baryonic. As a first order estimate, consider a Navarro, Frenk & White (1997) dark matter distribution with concentration parameters of Bullock et al. (2001), with a disk angular momentum parameter of $\lambda_d$~0.1 (characteristic for the five galaxies in Table 2, Bouché et al. 2007), and with a disk mass fraction (relative to the halo) of $m_d \geq \lambda_d$, set by the requirement of gravitational instability of the disk (Mo, Mao & White 1998). With these assumptions the range of maximum disk rotation velocities in Table 2 corresponds to halo circular velocities of 190±70 km/s and dark matter mass contribution of $10^{9.9\pm0.15}$ $M_\odot$ within 3 kpc of the center. This mass corresponds to 5-10% percent of the dynamical masses we are inferring in Table 2. The dark matter contributions to the *total* dynamical masses in Table 2 are larger. For the same assumptions as above the dark matter mass contributions within the disk/ring radii (~6kpc) are about 20% and within 10 kpc they are about 40% of the total dynamical masses.

A key question is how these early central bulges/central disks have formed. Are they the result of major dissipative mergers, as motivated by the CDM merger paradigm and many recent simulations (e.g. Steinmetz & Navarro 2002)? Or have they formed through internal, secular processes, as proposed for 'pseudo'-bulges in late type disks at z~0 (Kormendy & Kennicutt 2004)? While our current galaxy sample is too small to draw detailed and robust conclusions, we show below that the data nevertheless provide interesting new and perhaps unexpected insights into this important issue. Obviously a first hint is that none of the five galaxies in Table 2 exhibit dynamical or structural evidence for a major merger (Shapiro et al. 2008, Förster Schreiber et al. 2008b).

## 4.1 mass concentration parameter as a function of evolutionary state

In the following section we explore the dependence of the new mass concentration parameter on other properties of the galaxies. We find that $M_{dyn}(\leq0.4")/M_{dyn}(1.2")$ increases with evolutionary state/age.

For the five galaxies discussed in the last section the mass concentration parameter appears to increase with the [NII]/Hα flux ratio (upper left panel of Figure 10). Unless the [NII]/Hα ratio is strongly affected by a central AGN, [NII]/Hα is a proxy of the oxygen to hydrogen abundance ratio in the ionized interstellar medium (Pettini & Pagel 2004, Erb et al. 2006a). None of the five galaxies show kinematic evidence in our SINFONI data for a strong central AGN. The [NII]/Hα ratio in Table 3 refers to the value in the disk on scales of ~10 kpc, which should not be strongly affected by an AGN, even if a weaker or buried AGN is present in the center. Erb et al. (2006a) have studied the mass-metallicity relation at z~2. They found that [NII]/Hα scales with stellar mass, the 'mass-metallicity' relation. They also conclude that '….the z~2 mass-metallicity relation is driven by the increase in metallicity as the gas fraction decreases through star formation…..'. At a given mass the metallicity would thus appear to probe the evolutionary state of a galaxy.

The upper right panel of Figure 10 shows that in our z~2 SINS sample [NII]/Hα indeed is correlated with the 'star formation age', $M_*/SFR$ (Förster Schreiber et al. 2008a). This is the time required to form the current stellar mass at the current star formation rate. As a result the mass concentration parameter $M_{dyn}(\leq0.4")/M_{dyn}(1.2")$ also scales with star formation age. The star formation age, however, is necessarily less well determined than the [NII]/H$\alpha$ ratio because of the uncertainties in extinction and stellar mass. We thus prefer to show the correlation between $M_{dyn}(\leq0.4")/M_{dyn}(1.2")$ and [NII]/H$\alpha$ in the upper left panel of Figure10. In further support of the correlation between metallicity and evolutionary state three of our SINS galaxies (BX610, BzK6004 & BzK6397: Förster Schreiber et al. 2006, Buschkamp et al. 2008) exhibit a significant radial gradient of the [NII]/H$\alpha$ line ratio, implying that the central regions have a ~20% greater oxygen metallicity than the outer disks. The bulge regions of these three galaxies have super-solar metallicities (Buschkamp et al. 2008).

Obviously the statistical significance of a correlation based on five points is limited. Keeping this caveat in mind our data thus suggest a scenario where the fraction of the central bulge mass (including that of the surrounding central disk) appears to increase secularly over time scales of less then a few Gyrs.

### 4.2 turbulent and clumpy disks

The eight actively star-forming galaxies in Table 3 (~0.2-3 $M_\odot yr^{-1}kpc^{-2}$) all have a large component of random local gas motion, $\sigma_0$~45-90 km/s, in agreement with the earlier finding of Förster Schreiber et al. (2006). Given the spatial and spectral properties of the other ~50 galaxies in the SINS survey, this conclusion appears to hold generally for the entire population of z~2 star forming galaxies studied so far by integral field spectroscopy (Förster Schreiber et al. 2006, 2008a, Wright et al. 2007,

Law et al. 2007). The ratio $v_d/\sigma_0$ ranges from 2 to 5 in the seven large disk galaxies, as compared to 10-20 in z~0 disks (e.g. Dib, Bell & Burkert 2006). Yet smaller values of $v_d/\sigma_0$~1 are found for a number of the compact BX galaxies, such as BX502 (Table 3, Law et al. 2007 and in preparation, Förster Schreiber et al. 2008a). These latter systems appear to be gas rich and 'dispersion dominated'. The z~2 BX/s-BzK galaxies in the SINS survey are turbulent and geometrically thick ($h_z$ ~ 1 kpc).

The high resolution Hα and H/K images of the BX/s-BzK galaxies, as well as the rest-frame UV-continuum images in most other high redshift star-forming galaxies are typically dominated by a modest number (5-10) of bright, compact clumps (Cowie, Hu & Songaila 1995, van den Bergh et al. 1996, Elmegreen & Elmegreen 2005, 2006, Elmegreen et al. 2007, Förster Schreiber et al. 2008b). In many cases these giant star-forming complexes make up a significant fraction (~20-40%) of the mass of the disk (Elmegreen & Elmegreen 2005, Elmegreen et al. 2007, Genzel et al. 2006). The SINFONI AO data sets of BX482 and BzK15504 as well as the NIC2/ACS images of BX482, BX389, BX610 and MD41 have sufficiently high resolution for determining the intrinsic clump sizes. Their FWHM diameters, $L_c$, range from 0.15'' to 0.45'' (Figure 9). In BX482 (where both NIC2 and Hα data are available) stellar and gas clump sizes are comparable. Elmegreen and Elmegreen (2005, 2006) find similar results for the 'clump cluster'/'chain' galaxies in the HUDF, on the basis of photometric analysis of rest-frame UV/ACS images. In their study of 10 clump-cluster galaxies at $<z_{phot}>$~2.3 Elmegreen & Elmegreen (2005) infer an average clump size of $L_c$=0.22''. The HUDF galaxies are somewhat lower luminosity/mass analogs of the galaxies we are studying here ($<M_{gal}>$~$10^{10.7} M_\odot$, $<SFR_{gal}>$~20 $M_\odot yr^{-1}$, $<v_d>$~150 km/s, $<M_c>$~$10^{8.8} M_\odot$: Elmegreen & Elmegreen 2005). These galaxies appear to dominate the z>1 cosmic star formation in the HUDF (Elmegreen et al. 2007).

The Jeans length in a gravitationally unstable gas disk with $Q_{gas,Toomre} \leq 1$ is

$$L_{Jeans} = \frac{\pi}{\sqrt{2}}\left(\frac{\sigma_0}{v_d}\right) R_r Q_{Toomre} \qquad (2),$$

which yields $<L_{Jeans}> \sim 2.5$ kpc or 0.3" for the parameters appropriate for the galaxies in Tables 2 and 3. The observed sizes of the giant Hα clumps/star forming complexes in the clumpy z~2 galaxies are consistent with the Jeans lengths inferred from the galaxies' kinematics properties. The Hα clumps thus may be initially close to virial equilibrium. Given their large sizes and the overall large surface densities of the galaxy disks (a few $10^2$ $M_\odot pc^{-2}$), as derived from our dynamical measurements, it is clear that the star formation complexes must be very massive (see also Elmegreen & Elmegreen 2005)

$$M_c = \frac{L_c^2 v_d^2}{4GR_d} \sim 10^{9.4 \pm 0.5} \; M_\odot \qquad (3).$$

These masses are obviously much larger than even the most massive HII regions in z~0 starburst galaxies ($10^{7...8}$ $M_\odot$).

In the standard picture of disk formation from gas accretion from the halo (e.g. Mo, Mao & White 1998) the radius of the initial star-forming disk is set by the characteristic angular momentum parameter $\lambda_h$ ($<\lambda_h>$~0.04, Bett et al. 2007) of the accreting gas, $R_d \sim \lambda_h R_{vir}/\sqrt{2}$, where $R_{vir}$ (~100-150 kpc, Förster Schreiber et al. 2006) is the virial radius of the dark matter halo in which the disk resides. This scenario is in good agreement with the observed sizes of the BX/s-BzK disks in the SINS sample provided that the baryonic angular momentum is conserved during the accretion and that the larger systems we are discussing here are associated with halos of fairly high ($\lambda_h \sim 0.1$) angular momentum parameters (Bouché et al. 2007). Such high baryonic angular momentum parameters may be the result of the specific formation paths of their halos (Sales et al. 2008, in preparation).

### 4.3 secular evolution is efficient at high redshift

In a smooth gaseous disk the time scale for radial secular evolution is given by the viscous drag time (Shakura & Sunyaev 1973, Silk & Norman 1981, Lin & Pringle 1987, Silk 2001)

$$t_{visc} = \frac{R^2}{\eta} \text{ for viscosity } \eta \quad (4),$$

such that for $\eta = \alpha h_z \sigma_0$ (Shakura & Sunyaev 1973) with z-scale height $h_z$, dimensionless parameter $\alpha \approx 1$ and disk rotation velocity $v_d = \left(R/h_z\right)\sigma_0$

$$t_{visc} = \frac{R^2}{\alpha h_z \sigma_0} = \frac{1}{\alpha}\left(\frac{v_d}{\sigma_0}\right)^2 \left(\frac{R}{v_d}\right) = \frac{1}{\alpha}\left(\frac{v_d}{\sigma_0}\right)^2 t_{dyn}(R) \quad (5).$$

Once fragmentation and star formation sets in ($Q_{gas} \leq 1$, eq.2), the secular evolution time scale is then given by the dynamical friction time scale of the clumps against the background of the disk/halo. Based on Chandrasekhar's formula (Noguchi 1999, Silk 2001, Immeli et al. 2004a, b, Binney & Tremaine 2008) the dynamical friction time scale is

$$t_{df} = \beta\left(\frac{R}{\lambda_{Jeans}}\right)^2 t_{dyn}(R) = \beta\left(\frac{v_d}{\sigma_0}\right)^2 t_{dyn}(R) \quad (6),$$

with an (uncertain) dimensionless factor $\beta \sim 0.3$ (see above references). Here $t_{dyn}$ is the dynamical time scale, $t_{dyn} = 2.4 \times 10^7 R_6/v_{250}$ yrs. Here $R_6$ is the disk radius in units of 6 kpc and $v_{250}$ is the disk circular velocity in units of 250 km/s. Other secular processes in a stellar disk, such as bar formation and transport, occur on a time scale $t_{bar} \sim 30\text{-}50$ $t_{dyn}(R_6)$ (Immeli et al. 2004 a,b, Bournaud & Combes 2002).

Disk galaxies at z~0 have $v_d/\sigma_0$~10-20 (e.g. Dib et al. 2006) and secular processes proceed on a time scale of several Gyrs or more (Kormendy & Kennicutt 2004). The z~2 star- forming galaxies we are studying here are clearly different. In these galaxies $v_d/\sigma_0$ ranges between 1 and 5. Compared to z~0, gaseous and stellar processes proceed faster by at least one order of magnitude or more,

$$t_{sec}(z \sim 2) \sim 10 - 30\ t_{dyn}(R) \sim 0.5\ \text{Gyr} \sim t_{dyn}(R_{vir}) \sim 0.2\ t_H(z \sim 2) \quad (7).$$

Here $t_{dyn}(R_{vir})$ is the dynamical time for crossing the virial radius of the dark matter halo. Exactly which secular process dominates depends on the stellar and gas fraction, whether $Q_{gas}$ is greater or less than $Q_{stars}$ (Immeli et al, 2004 a,b), and whether bar instabilities occur in a turbulent environment.

We propose that in the five cases we have presented in this paper, we are witnessing the rapid buildup of central bulges through such secular processes. Our observations are in excellent agreement with the numerical simulation work of Noguchi (1999), Immeli et al. (2004a,b) and Bournaud, Elmegreen & Elmegreen (2007), which we discuss further in section 4.5.

Secular evolution of disks has been previously considered for the origin of the so-called 'pseudo-bulges' of z~0 late type spiral galaxies (Kormendy & Kennicutt 2004), growing over a timescale that is comparable to the present Hubble time. Such a long time scale appears to be at variance with the uniformly old age of stars in the bulge of the Milky Way (Zoccali et al. 2003), suggesting formation by early merging rather than late disk instability. Our finding of a very short timescale for the disk instability in z~2 disks reopens this issue, making it quite plausible that the Galactic bulge resulted from fast disk instability some 10 Gyr ago. In support of this conjecture Melendez et al. (2008) have found that the α/Fe-element abundances of the old Galactic bulge are more or less identical to those of the thick disk, implying that these two components are coeval and have both formed rapidly.

### 4.4 large turbulent velocities driven by cold flows

It is not obvious what causes the large velocity dispersions and large implied scale heights of the high-z disks. Here it is important to realize that the rest-frame UV-/optical continuum data of our sample galaxies as well as the HUDF galaxies of Elmegreen & Elmegreen (2005, 2006) imply the same ~1 kpc z-scale heights in the stars as in the ionized gas traced by Hα. The large turbulence is probably a property of the entire star-forming gas and stellar layer, and not just a feature of the ionized gas. One obvious possibility is that these motions are the result of feedback from the intense star formation itself: by the combined actions of winds, supernova explosions and radiation transport (Efstathiou 2000, Silk 2001, Thompson, Quataert & Murray 2005). Theoretical estimates and extrapolations from local starburst galaxies (Monaco 2004, Dib, Bell & Burkert 2004) make a plausible case that the observed velocity dispersions (45-90 km/s) can be reached at the observed star formation surface densities in the BX/s-BzK galaxies. In this case one would expect that the $v_d/\sigma_0$ ratio scales inversely with the star formation rate surface density $\Sigma_{SFR}$. The bottom left panel of Figure 10 is a first attempt of testing this assertion. There is no obvious correlation between the observed values of $v_d/\sigma_0$ and $\Sigma_{SFR}$ among the 7 massive galaxies in Table 3. The turbulence in the compact, young and high star forming surface density system BX502 is much greater than in the other galaxies. Including this system may add some evidence that feedback may indeed be at work in the systems with the largest star formation surface densities. BX502 has a much lower mass than the other seven galaxies, and it is not yet clear how these compact dispersion limited systems relate to the massive large disks and whether they have experienced a recent merger (Law et al. 2007).

A second possibility is that the turbulence is driven by the accreting gas as it enters the forming disk. Recent high-resolution adaptive mesh simulations by Ocvirk, Pichon & Tessier (2008) clearly demonstrate that in the cold flow regime (Dekel & Birnboim 2006, Kereš et al. 2005) gas is brought in radially along the filaments of the cosmic web, from well outside the virial radius of the halo and all the way to the disk region. It would appear to be unavoidable that the rapidly inflowing ($v_{inflow}=\gamma v_d$, with a dimensionless number $\gamma\sim 1$-1.4) gas streamers and clumps (minor mergers?) embedded in the cold flows shock and stir up the forming protodisk (e.g. Abadi et al. 2003). There are no realistic simulations yet of the disk/flow interaction and shocks. The rate of specific energy gain is $\dot{E}_{acc}=\beta\gamma^2 v_d^2/t_{acc}(z)$, where $t_{acc}=<M_{gas}/(dM_{acc}/dt)>_z$ is the gas accretion time at z. If the rate of turbulent energy loss of the disk clumps is driven by their physical collision rate $\dot{E}_{coll}=-\sigma_0^2/t_{coll}$, the velocity dispersion is given by (Förster Schreiber et al. 2006)

$$\sigma_0 \sim \beta\gamma^2 v_d \sqrt{\left(t_{coll}/t_{acc}\right)} \qquad (8).$$

For the parameters of the observed disks with 5-10 virialized clumps of radius ~1 kpc, the collisional time scale is close to the dynamical time scale of the ring/disk, $t_{coll}=\varepsilon t_{dyn}$ with $\varepsilon\sim 1$. From a statistical analysis of the Millenium dark matter simulation (Springel et al. 2005), Genel et al. (2008, see also Birnboim, Dekel & Neistein 2007) derive a fitting formula for the average accretion time scale of dark matter as a function of z(t) and halo mass $M_h$,

$$<t_{acc}(z,M_h)> \ = \ 2.27 \times 10^9 \left(\frac{1+z}{3.2}\right)^{-2.2} \left(\frac{M_h}{2 \times 10^{12} M_\odot}\right)^{-0.08} \ (\text{yrs}) \qquad (9).$$

For the cosmic baryon fraction of ~0.2 the baryonic accretion rates implied by equation (9) would also be sufficient to explain sustained star formation rates of >100$M_\odot$yr$^{-1}$ in the massive BX/s-BzK galaxies in this paper. Equations 8 and 9 then yield

$$\left(\frac{v_d}{\sigma_0}\right) \ \sim \ 9.7 \ \beta^{-1}\gamma^{-2}\left(v_{250}R_6^{-1}\varepsilon^{-1}\right)^{1/2}\left[(1+z)/3.2\right]^{-1.1} M_{2e12}^{-0.04} \qquad (10),$$

where $M_{2e12}$ is the halo mass in units of $2 \times 10^{12}$ $M_\odot$. These numbers are appropriate for the massive BX/s-BzK galaxies we are considering here. The simple estimate in equation (10) suggests that the conversion of accretion energy may account for the observed turbulent motions if the product $\beta \gamma^2 \varepsilon^{1/2}$ exceeds ~2. Equation (10) also indicates that over 2-3 Gyrs of cosmic evolution (from z~4 to z~2) the average $v_d/\sigma$ ratio would be expected to increase by a factor of 1.75, from the cosmological decrease of the accretion rate. A still larger decrease in the accretion rate would be predicted once the halos grow above the hot/cold flow boundary (Dekel & Birnboim 2006, Ocvirk et al. 2008) and if AGN feedback is effective. Stirring of the disk turbulence by the accretion flows would thus naturally predict that an early phase of rapid accretion leading to clumpy, thick protodisks and secular bulge formation is followed a few Gyr time scale later by a less violent re-growth of disks, perhaps related to the formation of thin disks.

The lower right panel in Figure 10 is an attempt of exploring whether there is a trend of $v_d/\sigma_0$ with evolutionary state, again as above in form of the [NII]/Hα ratio. With the exception of BX482 there may indeed be a trend for the more mature sources to be also the ones with the smallest relative fraction of turbulent motion (largest $v_d/\sigma_0$). As with the other trends discussed in this section, a larger sample of well resolved kinematic data sets will be required to make more definite statements.

The scenario we are discussing relies on the assumption that the accreting material is highly gas-rich. A simple check of this assertion is possible if one assumes that below a certain critical halo mass, supernova feedback prevents early star formation to convert a large fraction of the baryonic gas into stars (Dekel &Silk 1986, Efstathiou 2000). Following Dekel & Silk (1986) and Dekel & Woo (2003) this critical baryonic mass is given by

$$M_{b,crit} = 2.0 x 10^{10} \left( \frac{v_{d,crit}}{100 \text{ km/s}} \right)^3 \left( \frac{1+z}{3.25} \right)^{-2/3} \quad (11),$$

where we have again assumed a baryon fraction in the disk of b~0.2. Halos with masses below $M_{b,crit}$/b form stars inefficiently and should be gas rich. This critical mass is a factor 3.5 to10 smaller than the dynamical/stellar masses of the five galaxies in Table 2. Given that the galaxies we have observed appear not to be undergoing major mergers, any minor merger accretion would have to have a mass ratio >3:1. Such minor mergers would be at or below $M_{crit}$, and thus would be very gas rich according to equation (11). Here we presume that in the limit of very minor mergers (mass ratios >>10:1) smooth gas accretion and a series of minor mergers are virtually identical.

We conclude from this section that if the trends suggested by the present data are confirmed, the level of turbulence in the high-z disks may be strongly influenced by (the evolution of) the accretion rate, as well as stellar feedback in the most intensely star forming systems. The large observed turbulence of the high-z disks may be plausibly understood with current estimates of 'cold flow' gas accretion rates based on dark matter simulations.

**4.5 Comparison to simulations**

Stimulated initially by the clumpy appearance of high-z galaxies on HST rest-frame UV-images ('chain' galaxies: Cowie, Hu & Songaila 1995, van den Bergh et al. 1996) and later by the work of Elmegreen & Elmegreen (2005, 2006), several groups have carried out numerical simulations of the evolution of gas-rich, fragmentation unstable, 'clumpy' galaxies with star formation recipes (Noguchi 1999, Semelin & Combes 2002, Immeli et al. 2004a,b, Bournaud et al. 2007). These simulations vary in technical approach (N-body simulations vs. hydro-grid code) and initial assumptions. For instance, Bournaud et al. (2007) adopt an ab initio unstable disk while Semelin & Combes (2002) start with hot gas that then becomes gravitationally unstable after cooling to the molecular phase. Noguchi (1999) and Immeli et al. (2004a,b) let the disk to grow to instability from rapid gas inflow along the polar axis. The latter authors start with a pure gas disk in a dark matter halo. Bournaud et al. introduce, in addition, an initial stellar component making up 50% of the mass. While Bournaud et al. assume that the initial stellar disk is already geometrically thick but do not include stellar (supernova) feedback, Immeli et al. consider supernova feedback with a two-phase interstellar medium. Keeping in mind these differences as well as the simplifications that none of these simulations begin from self-consistent cosmological initial conditions, all these studies find essentially the same salient features;

- rapidly forming, very gas-rich disks will become violently and globally unstable to fragmentation into giant star forming clumps if the Toomre instability parameter in the gas layer suddenly becomes $Q_g \leq 1$ across much of the entire disk. This requires that the (cold) gas accretion rate is high (>100 $M_\odot yr^{-1}$) and happens on a time scale comparable to the orbital time scale of the initial disk (~0.5 Gyrs in the simulations of Noguchi 1999 and Immeli et al. 2004a,b);

- once fragmentation sets in, the disk evolves rapidly during a short-lived (0.4-1 Gyr) 'clump phase'. During this phase star formation proceeds rapidly in the giant clumps, with a star formation rate comparable to the gas accretion rate. The strong dynamical interactions heat the gas and stellar components with a resulting z-scale height of ~ 1 kpc;

- as a result of efficient dynamical friction of the clumps against the background of the rest of the disk, the clumps spiral into the center and form a central bulge and a surrounding smooth exponential stellar disk. Clump-clump collisions dissipate the large turbulent motions.

Samland & Gerhard (2003) consider a case with a much lower accretion rate (~20-40 $M_\odot yr^{-1}$). In this model, the disk grows from inside out, with a bulge forming first from low angular momentum gas inflow. The surrounding disk is quite thick and the most active star formation occurs in a ring that moves over time outwards from the center. The reason for outward propagation is that the gas infall can no longer compensate for the gas consumption by star formation in the center. This in turn is caused by the decrease in gas accretion to the central regions due to increasing specific angular momentum of the infalling material. The ring grows in radius until eventually the disk becomes unstable and forms a bar.

One prediction of these simulations and the generic hypothesis of clump evolution from smooth accretion of gas from the halo is that the clumps would plausibly have similar metallicities while the interclump gas would be expected to have much lower metallicities. This is indeed consistent with our current data, to within the limitations of signal-to-noise ratio and resolution. With the exception of radial [NII]/Hα gradients mentioned above there are no indications for significant [NII]/Hα variations from clump to clump in the disk regions of the five galaxies of our main sample, nor in the three additional larger systems in Table 3. The signal to noise ratio is not yet sufficient to explore the interclump metallicity.

It is remarkable how well the observed properties of the young disks we are studying in this paper, the simple 'back of the envelope' considerations presented above and the more detailed simulations appear to be consistent with each other. The interesting consequence and prediction of these simulations is that these clumpy, turbulent disk galaxies are going through a transitory 'fragmentation' phase at the very beginning of their evolution.

### 4.6 Comments on asymmetries

In this paper and in our previous papers (e.g. Förster Schreiber et al. 2006, Genzel et al. 2006) we have emphasized the surprisingly symmetric properties of the kinematics of the UV-/optically selected massive star forming galaxies we have observed. This is an obvious (and intended) simplification. Asymmetries of the brightness and velocity distributions are present in essentially all the SINS data sets (this paper, Förster Schreiber et al. 2006, 2008, Genzel et al. 2006). As discussed in the section on BX482 (3.1) local variations in extinction, excitation and local current star formation events may account for some of these asymmetries even if the underlying kinematics and potential are intrinsically symmetric. However, there are clearly are also intrinsic variations in the kinematic properties that can be seen from asymmetric rotation curves (e.g. Figures 7 & 8) and from residuals in the velocity distributions (e.g. Figures 2 & 4) after subtraction of a first order rotation model (see Genzel et al. 2006, Cresci et al. 2008, and Förster Schreiber et al. 2008 for more examples). These features may be a natural consequence of asymmetries in the underlying potential and gas accretion processes. Lopsided rotation curves as observed in a number of the SINS galaxies have also been seen in the HI kinematics of some z~0 galaxies (Swaters et al 1999) and are ascribed there to long-lived intrinsic m=1 modes in the underlying potential and to interactions with nearby galaxies. At high redshift strong asymmetries would be naturally be expected to be caused by the non-isotropic and filamentary accretion, by the large radial velocities of the accreting gas and by the minor merger activity (e.g. Ocvirk et al. 2008). In the framework of the gas rich, clumpy galaxies very significant local perturbations from a smooth rotation curves are expected as a result of the large clump masses (Immeli et al. 2003a,b, Bournaud et al. 2007).

## 4.7 Consequences for galaxy evolution

We have presented evidence that massive bulges and spheroids (with masses up to at least $5 \times 10^{10}$ M$_\odot$) may have formed on a time scale of 1-3 Gyrs through secular evolution from gas-rich, turbulent disks. The upper end of this mass range would be consistent with bulge masses for z~0 Sa bulges. These turbulent disks grow from steady and rapid gas accretion (including minor mergers) from the dark matter halos and fragment into very large star formation clumps. Unless major mergers happen during this phase, it would appear that most galaxies experience such a 'clump'-phase that is an inevitable consequence of the combination of rapid continuous gas accretion, large gas fraction and large turbulence. This conclusion is consistent with the large fraction of such clumpy galaxies in the HUDF (Elmegreen et al. 2007) and the strong correlation between stellar mass and star formation in the z~2 UV- and optically selected galaxies (Erb et al. 2006c, Daddi et al.2007). In addition to rapid bulge buildup this process would also account rather easily for the fact that there are no descendent massive, thick disks at z~0 (Förster Schreiber et al. 2006). Estimates based on the Millenium simulation suggest that the major merger rate is too low to destroy all $M_*$~$10^{11}$ $M_\odot$ galaxies between z~ 2 and 0 (Genel et al. 2008, Fakhouri & Ma 2008, Conroy et al. 2008). It is tempting to speculate that the thick, old stellar disks seen in the Milky Way and nearby galaxies (e.g. Freeman & Bland-Hawthorn 2002, Yoachim & Dalcanton 2006) are the remnants of this z~2-3 phase of transient evolution of clumpy, turbulent disks fed by rapid cold flows.

Efficient secular evolution may also help to explain the remarkably efficient and rapid star formation in the z~2 star-forming galaxy population (Förster Schreiber et al. 2006, Genzel et al. 2006, Daddi et al. 2007). Davé (2007) has introduced the star formation activity parameter

$$\alpha_{SF}(z) = \left( \frac{M_*}{SFR \cdot (t_H(z) - 1 \text{ Gyr})} \right) \qquad (12),$$

where $\alpha_{SF}$ is equal to unity if the stellar formation time scale $M_*/SFR$ is just equal to the available cosmic (Hubble) time at z, taking into account that most of the star formation occurred after re-ionization (< 1Gyr). Figure 11 shows that most of the observed z~2 star-forming s-BzK and BX galaxies, including the majority of the SINS survey galaxies (Förster Schreiber et al. 2008b) and the five galaxies in Table 3 with a determination of $M_*$, appear to have formed their stars on a time scale a few times faster than the available Hubble time (see also Daddi et al. 2007). This is in disagreement with the simulations for which $\alpha_{SF}$~1 (Davé 2007, Kitzbichler & White 2007).

Secular evolution together with efficient star formation at high gas surface densities may explain why the star formation activity parameter $\alpha_{SF}$ is significantly less than one at high redshift. For a Schmidt-Kennicutt star formation recipe, the star formation rate surface density scales approximately inversely proportional to dynamical time scale (Kennicutt 1998, Bouché et al. 2007). If the gas in the initial disks fragments into compact dense clumps and at the same time the overall disk contracts as a result of the dynamical friction process, the star formation activity parameter would naturally decrease by a factor of several. Further, Bouché et al. (2007) have presented evidence that the star formation efficiency per dynamical time scale in high density star-forming galaxies, such as the z~2 BX/s-BzKs, is about 0.07, or 3 times larger than that estimated for normal z~0 spirals (Kennicutt 1998). For this purpose gas masses were inferred from CO rotational line luminosities calibrated at the low gas surface density end by the normal z~0 star forming galaxies from Kennicutt (1998). On the high end the calibration is by z~0 ULIRGs (Kennicutt 1998) and z~1-3 SMGs (Tacconi et al. 2006, 2008). For the normal galaxies Bouché et al.

(2007) used a Milky Way CO-to-$H_2$ conversion factor (as did Kennicutt 1998), while for ULIRGs and SMGs Bouché et al. applied a CO-to-$H_2$ conversion factor ~0.2 times that of the Milky Way (see Tacconi et al. 2008 for a more detailed discussion). For the gas surface densities derived in this way for the BX/s-BzK galaxies the Bouché et al. (2007) fit gives the following expression for the $\alpha_{SF}$ parameter

$$\alpha_{SF}(z \sim 2) = 0.24 \left( \frac{\Sigma_{gas}/\tau_{dyn}}{20 \text{ M}_\odot \text{yr}^{-1}\text{kpc}^{-2}} \right)^{-0.14} R_{6kpc} v^{-1}_{250km/s} \qquad (13).$$

Since the above estimate comes mainly from dynamical measurements and gas fractions calibrated on CO observations from gas-rich, z~0-3 mergers (Tacconi et al. 2008), the above expression would appear to be relatively immune to changes in the assumed initial mass function. With the Bouché et al. (2007) calibration of the Kennicutt-Schmidt relation the gas fractions in the five galaxies we have studied range between 0.16 and 0.31 of the dynamical masses (Table 2) and the gas exhaustion time scales, $M_{gas}$/SFR(H$\alpha$), are ~$2\times10^8$ yrs, significantly smaller than the SED ages or the stellar buildup ages, $M_*$/SFR(H$\alpha$). As it is highly unlikely that all BX/BzK galaxies are about to run out of gas and will soon stop forming stars, this discrepancy argues for a continuous gas supply (Erb 2008), and thus is consistent with the continuous gas accretion model. However, the gas masses in Table 2 are obviously based on a somewhat uncertain, indirect method. Direct measurements of gas masses from CO rotational line luminosities would be highly desirable. First CO observations in a few BX/BzK galaxies have resulted in detections in two cases (possibly implying much larger gas masses than in Table 2: Daddi et al. 2008) and upper limits in three other cases (Tacconi et al. 2008). These initial observations do not yet provide a clear-cut test of whether the Bouché et al. (2007) calibration is appropriate, especially when taking into account the uncertainty in the CO-

luminosity-to-$H_2$-gas mass conversion factor (see Tacconi et al. 2008 for a discussion).

The combination of efficient secular evolution and higher efficiency star formation may thus be able to account for the observed rapid buildup of stellar mass at high redshift. It needs to be explored whether secular evolution may also be able to influence the buildup of the central massive black holes and account for the black hole mass to bulge mass (or velocity dispersion) relation.

## *5. Conclusions*

We have presented high quality SINFONI/VLT integral field spectroscopy of five large $M \sim 10^{11} M_\odot$ disk galaxies at $z \sim 2$. Aided in part by laser guide star adaptive optics, our data constrain the galaxy dynamics on scales of $\leq 3$ kpc and allow us to detect and estimate the dynamical mass of the central bulge/inner disk, in addition to that of the outer disk.

We find evidence for a substantial central bulge and/or central disk in four of the five cases, whose mass fraction relative to that of the disk appears to scale with the [NII]/H$\alpha$ ratio and star formation age of the galaxies.

Interpreting these observations in terms of an evolutionary sequence, we propose that massive bulges/central disks at $z \sim 2$-3 can grow on a time scale of $\leq 1$ Gyr through secular evolution of early turbulent, gas-rich disks. Numerical simulations and simple analytical estimates suggest that clumpy, turbulent disks may be a transient phase in the early evolution of most, if not all, galaxies that are going through a phase of rapid 'cold' accretion from the surrounding halos. To a significant extent the large turbulence may be stirred up by the kinetic energy of the accreting gas itself. Precisely because of the large turbulence, internal secular disk evolution then proceeds on a time scale ≤1 Gyr, an order of magnitude more rapidly than at z~0. After the clumpy, turbulent phase is ended, the product may be a (massive) central bulge surrounded by a remnant thick exponential disk whose z~0 relic is the old thick disk component in many nearby galaxies. As a result of the cosmological evolution of accretion rates combined with the ceasing of rapid cold flows once halos grow above ~$10^{11.5-12}$ $M_\odot$, accretion rates strongly decrease with decreasing redshift such that the disk turbulence subsides and maturing thin disks grow at z≤1.

These secular effects, along with more efficient star formation at high gas surface densities may also help to account for the ~1Gyr time scales for the buildup of the stellar component that appears to be typical at z~2. One is left to wonder whether the same secular processes also help build up central black holes in this epoch of maximum QSO activity.

*Acknowledgements: We thank the staff of Paranal Observatory for their support. This work would have been impossible without the dedicated work of the ESO LGSF team at ESO, MPE and MPIA Heidelberg. We are grateful to Norm Murray, Eve Ostriker and Chris McKee for enlightening discussions on star formation processes and to Avishai Dekel and an anonymous referee for valuable comments on the paper.*

**Table 1. Observing Log**

| Galaxy | band/pixel scale | mode | FWHM resolution (arcseconds) | integration time, observing date | reference |
|---|---|---|---|---|---|
| Q2346-BX482 (z=2.258) | K 0.05"x0.1" | LGSF | 0.2" | 3 h, Oct 07 | Erb et al. 06b,c |
|  | K 0.125"x0.25" | seeing limited | 0.5" | 8.8 h, Aug,Sep,Oct 05, Aug 07 |  |
|  | H 0.125"x0.25" | seeing limited | 0.6" | 4 h, Aug, Nov 06 |  |
| BzK6004-3482 (z=2.387) | K 0.125"x0.25" | seeing limited (excellent seeing) | 0.45" | 5h, Mar 06 | Kong et al. 06 |
|  | K 0.125"x0.25" | LGSF | 0.45" | 5h, Mar 07 |  |
| Q2343-BX389 (z=2.174) | K 0.125"x0.25" | seeing limited | 0.5" | 4h, Oct06 | Erb et al. 06b,c Förster Schreiber et al. 06 |
| Q2343-BX610 (z=2.211) | K 0.125"x0.25" | seeing limited | 0.5" | 6 h, June 05, Aug/Sep 05 | Erb et al. 06b,c Förster Schreiber et al. 06 |
| SSA22-MD41 (z=2.172) | K 0.125"x0.25" | seeing limtd | 0.5" | 8 h, Nov 04, June 05 | Erb et al. 06b,c Förster Schreiber et al. 06 |

# Table 2. Derived properties of the z~2 star-forming galaxies: part 1

| Parameter | MD41 (z=2.17) | BX482 (z=2.26) | BX389 (z=2.17) | BX610 (z=2.21) | BzK6004 (z=2.39) |
|---|---|---|---|---|---|
| $R_{ring}$ (kpc) | 5.9 (0.6) | 7.0 (0.8) | 4.4 (0.5) | 4.4 (0.5) | 6.9 (0.8) |
| inclination (degrees) | 71 (5) | 65 (7) | 80 (5) | 33 (5) | 35 (8) |
| SFR[1] ($M_\odot yr^{-1}$) | 75 | 140 | 150 | 210 | 160 |
| $M_{dyn}(\leq 20 kpc)$[2] ($M_\odot$) | 0.7 (0.15) $10^{11}$ | 1.4 (0.2) $10^{11}$ | 1.4 (0.2) $10^{11}$ | 1.5 (0.5) $10^{11}$ [3] | 1.9 (0.5) $10^{11}$ |
| $M_*$ ($M_\odot$) | 0.17-0.36 $10^{11}$ | 0.43-0.89 $10^{11}$ | 0.7 (0.2) $10^{11}$ | 1.7 (0.3) $10^{11}$ | 5.8 (2) $10^{11}$ |
| $M_{gas}$[4] ($M_\odot$) | ~0.2 $10^{11}$ | ~0.3 $10^{11}$ | ~0.2 $10^{11}$ | ~0.3 $10^{11}$ | ~0.3 $10^{11}$ |
| $M_{gas}/M_{dyn}$ | ~0.3 | ~0.2 | ~0.2 | ~0.1 | ~0.1 |
| $M_{dyn}(\leq 0.4")/M_{dyn}(1.2")$[5] | ≤ 0.15[5] (3σ) | 0.205 (0.03) | 0.39 (0.08) | 0.39 (0.08) | 0.37 (0.04) |
| comments on fit | exponential disk model produces too centrally concentrated light distribution | exponential disk model produces too centrally concentrated light distribution | exponential disk compatible with data <1" but produces too much light at >1" | exponential disk compatible with available data to ~1.2" | exponential disk does not fit central velocity distribution and outer rotation curve |

[1] corrected for extinction and Calzetti (2001) recipe, and for Chabrier IMF, uncertainties ≥50%

[2] for the ring/flat disk plus central bulge (HWHM in radius 0.3") model

[3] the best fitting exponential disk model has $2.3 \times 10^{11} M_\odot$ within 10 kpc

[4] from the 'Kennicutt-Schmidt' star formation recipe of Bouché et al. (2007): $M_{gas} = 3.67 \times 10^8 \{SFR (M_\odot yr^{-1})\}^{0.58} \{R_d (kpc)\}^{0.83}$ ($M_\odot$)

[5] ratio of dynamical mass within 0.4" (3kpc) and 1.2" (10 kpc), for the specific Gaussian ring/disk, plus central Gaussian bulge model. Error bars are the 1σ statistical errors for the fitting of that ratio, for all other parameters held constant

# Table 3. derived properties of BX/s-BzK galaxies: part 2

|  | BX502 | MD41 | BX482 | BX389 | ZC782941 | BzK15504 | BX610 | BzK6004 |
|---|---|---|---|---|---|---|---|---|
| $v_d$ (km/s) | 75 (25) | 175 (30) | 235 (40) | 265 (40) | 285 (30) | 230 (30) | 290 (70) | 255 (40) |
| $R_d$ (kpc) | 1.7 (0.5) | 5.9 (0.6) | 7 (0.8) | 4.4 (0.5) | 3.6 (1) | 4.5 (0.8) | 4.4 (0.5) | 6.9 (0.8) |
| [NII]/H$\alpha$ | 0.073 (0.026) | 0.08 (0.02) | 0.11 (0.026) | 0.21 (0.035) | 0.24 (0.025) | 0.36 (0.04) | 0.38 (0.035) | 0.42 (0.035) |
| M*/SFR(H$\alpha$) (Gyr)[1] | 0.04 | ---- | ----- | 0.5 | ----- | 0.8 | 0.8 | 3.6 |
| t* (SED) (Gyr) | 0.23 (0.15) | ---- | ----- | 2.7 (2) | ----- | 1.6 (+0,-1) | 2.7 (2) | 2.5 (+0,-1.2) |
| EW (H$\alpha$) (Å) | 2200 (550) | >130 | ---- | 300 (75) | 150 (38) | 125 (30) | 145 (36) | 68 (17) |
| $\Sigma_{SFR}$[1] ($M_\odot$ yr$^-$kpc$^{-2}$) | 3.6 | 0.35 | 0.57 | 0.6 | 1.5 | 1.2 | 1.2 | 0.6 |
| $\sigma_0$ (km/s) | 77 (10) | 79 (10) | 55 (5) | 87 (10) | 88 (15) | 45 (5) | 60 (7) | 60 (7) |
| $v_d/\sigma_0$ | 1 (0.4) | 2.2 (0.7) | 4.3 (0.8) | 2.9 (0.7) | 3.2 (0.6) | 5.1 (0.7) | 4.8 (0.8) | 4.3 (1.5) |

[1] uncertainties ≥50%

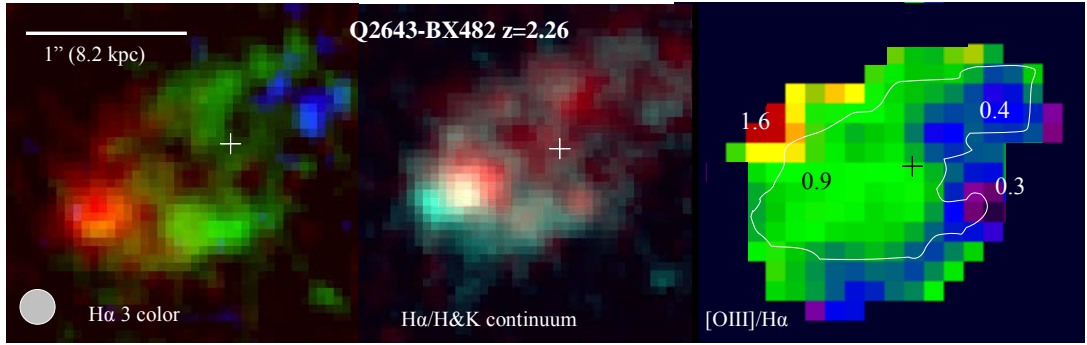

Figure 1. Hα, [OIII] and H&K continuum observations of Q2343-BX482 (z=2.26, Erb et al. 2006a-c, Förster Schreiber et al. 2006), at a resolution of 0.2" FWHM. Left: 3-color composite of blue-shifted, central and red-shifted Hα line emission, from SINFONI LGSF data at 0.2" FWHM (shaded circle). Middle: Integrated SINFONI Hα emission (light blue) superposed on NIC2 $H_{160}$ continuum (red) from Förster Schreiber et al. (2008b), with a resolution comparable to the Hα data. The images were aligned on the brightest spot in the south-east part of the ring whose K-band continuum is also detected in the SINFONI cube. The white cross marks the dynamical center of the object as determined by our modeling. Right: Observed [OIII] 5007/Hα flux ratio from 0.125"x0.25" pixel, seeing-limited observations and smoothed with a 0.6" kernel.

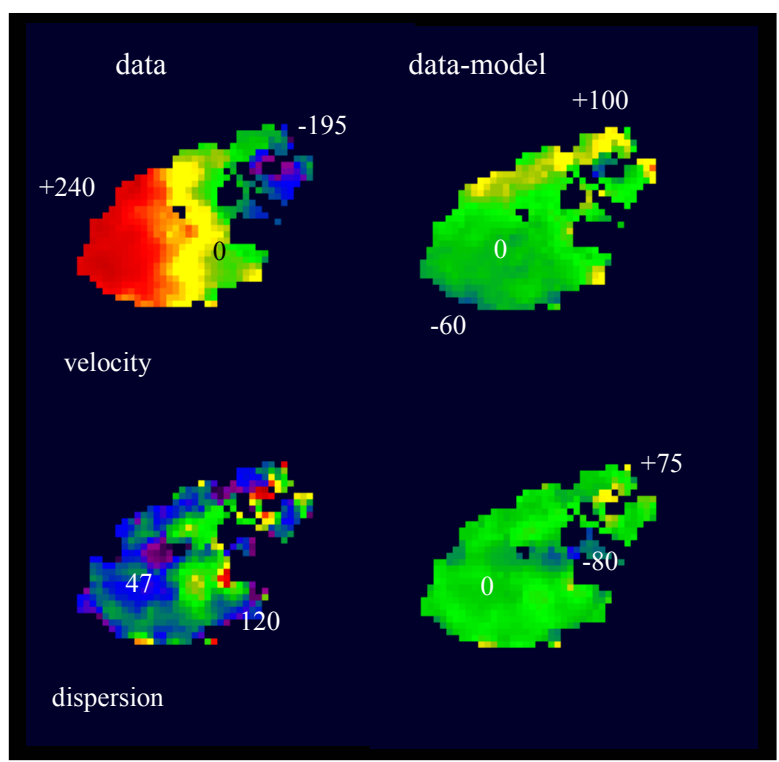

Figure 2: Two dimensional Hα velocity field and kinematic modeling of BX482. Left column: velocity centroid map (top) and velocity dispersion map (bottom). Right column: residual map (data minus model) of velocity (top) and velocity dispersion (bottom). The model Hα emission is the best fitting rotating ring plus bulge, as described in the text. Labels are in km/s and denote the extreme and zero color scales.

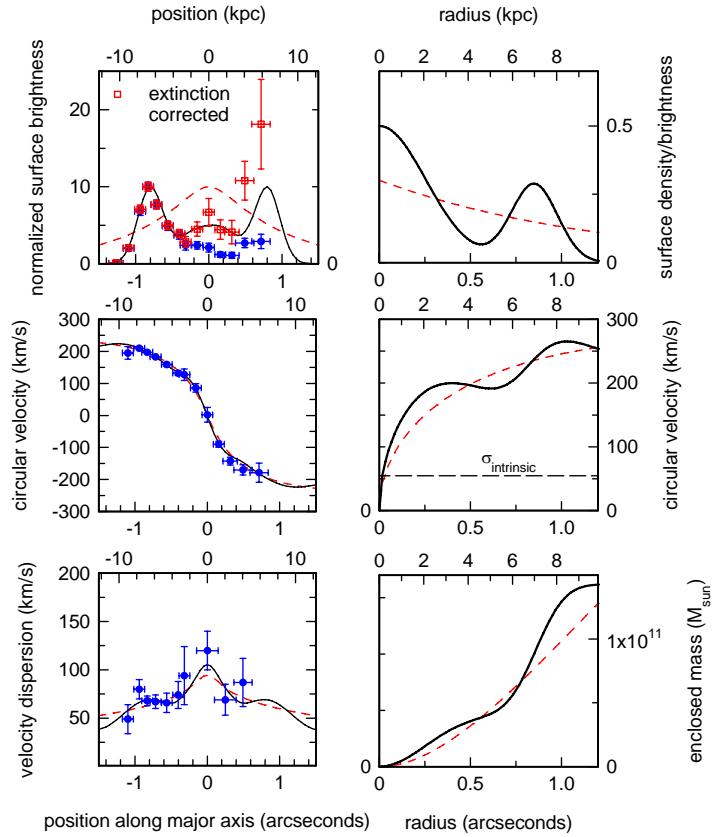

Figure 3: Hα major axis cuts and kinematic modeling of BX482. Left column: Kinematic major axis cuts in projected Hα flux (top), projected velocity (middle) and projected velocity dispersion (bottom), along p.a.-65$^0$. Blue filled circles are the measurements. Vertical-axis bars are the ±1σ errors, horizontal bars denote the aperture size. The black continuous lines represent the best fitting ring plus bulge model, and the red dotted lines the best fitting exponential disk model. The red squares in the upper left panel denote the Hα fluxes corrected for screen extinction (see text). Right column: Radial cuts of the best fitting, ring + bulge (continuous black) and exponential (red dotted) model distributions for the enclosed mass (bottom), intrinsic rotation curve (middle) and mass/brightness distribution (top). The dotted horizontal line in the middle panel marks the level of the intrinsic velocity dispersion in the ring, as derived from the residual velocity distribution.

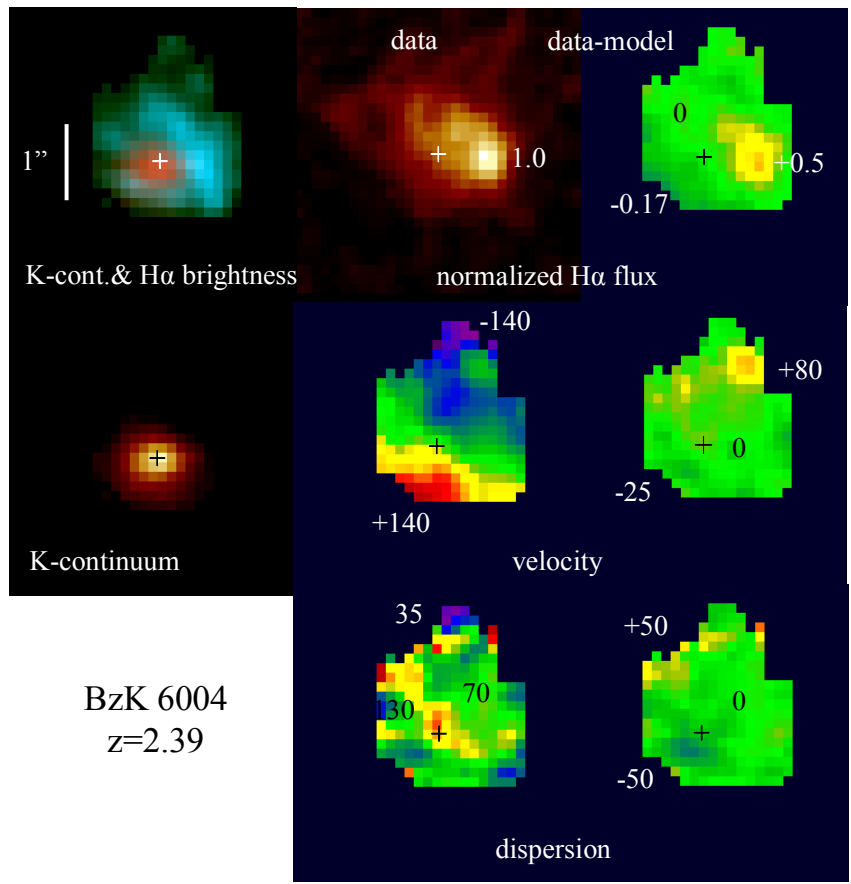

Figure 4. SINFONI LGSF Hα data set of D3a6004-3482 (z=2.387, Kong et al. 2006), at a resolution of 0.45" FWHM. Left column: Superposition of K-band continuum (red) and Hα surface brightness density (peak surface brightness per wavelength interval, blue) (top) and K-band continuum (middle). Second from left: normalized integrated Hα emission (top), velocity map (middle) and velocity dispersion map

(bottom). Third from left: residual map (data minus model) of Hα integrated emission (top), velocity (middle) and velocity dispersion (bottom). Numbers in the bottom four panels are in km/s and denote the extreme and zero color scales.

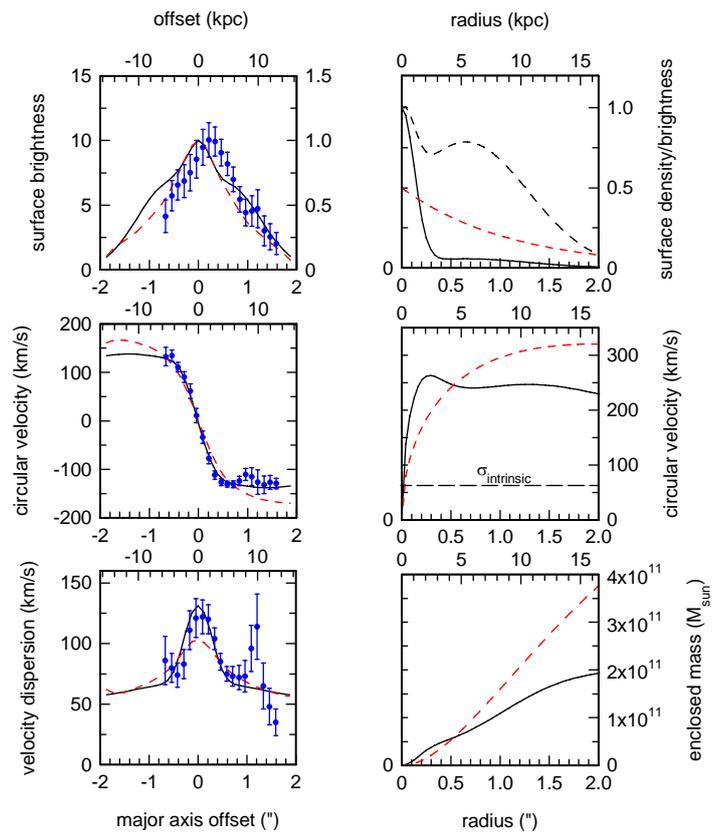

Figure 5. Hα major axis cuts and kinematic modeling of BzK6004. Symbols and markers are the same as in Figure 3 for BX482.

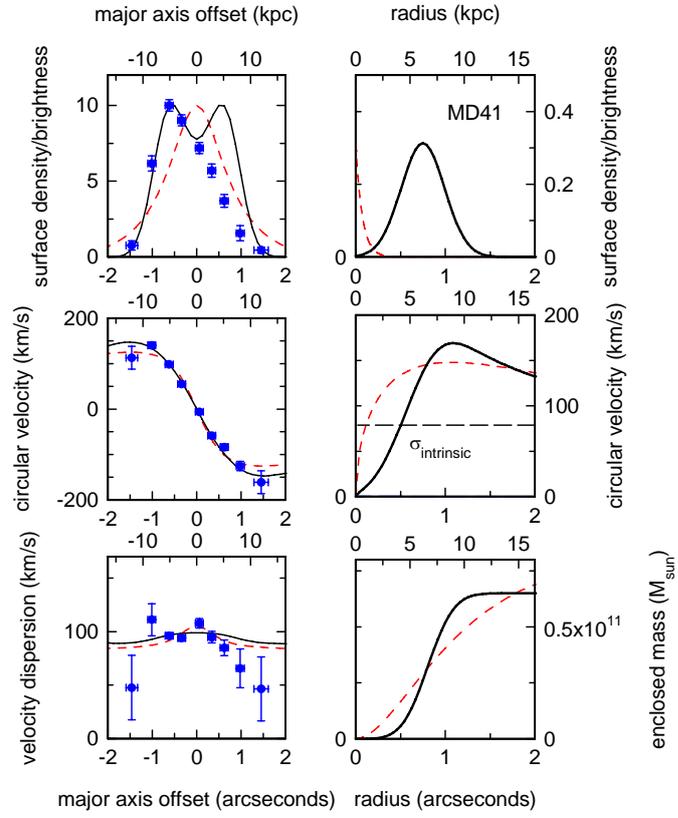

Figure 6. Hα major axis cuts and kinematic modeling of SSA22-MD41 (z=2.172) at a resolution of ~0.5" FWHM (see also Förster Schreiber et al. 2006). Symbols and markers are the same as in Figure 3.

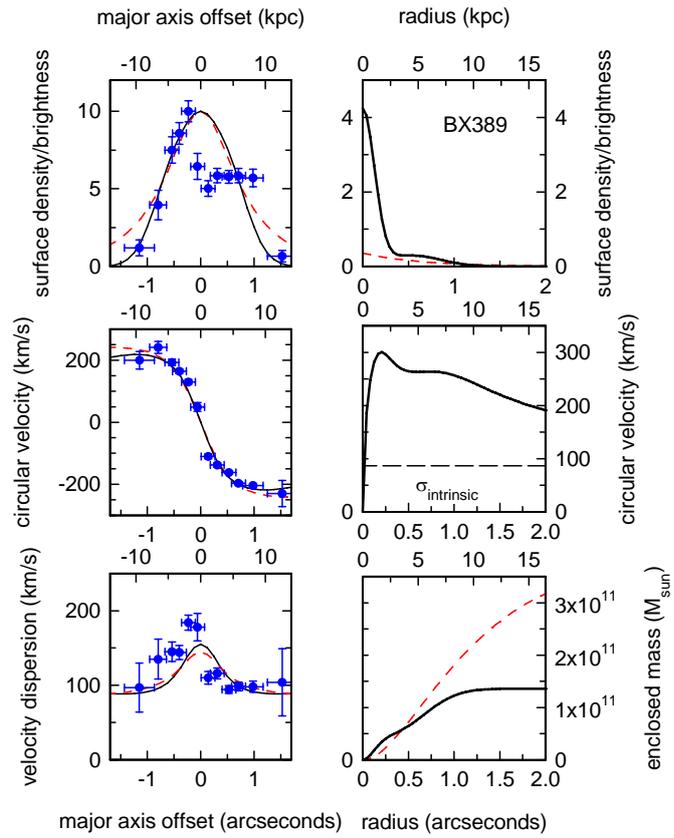

Figure 7. Hα major axis cuts and kinematic modeling of Q2343-BX389 (z=2.174), at a resolution of ~0.5" FWHM (see also Förster Schreiber et al. 2006). Symbols and markers are the same as in Figure 3.

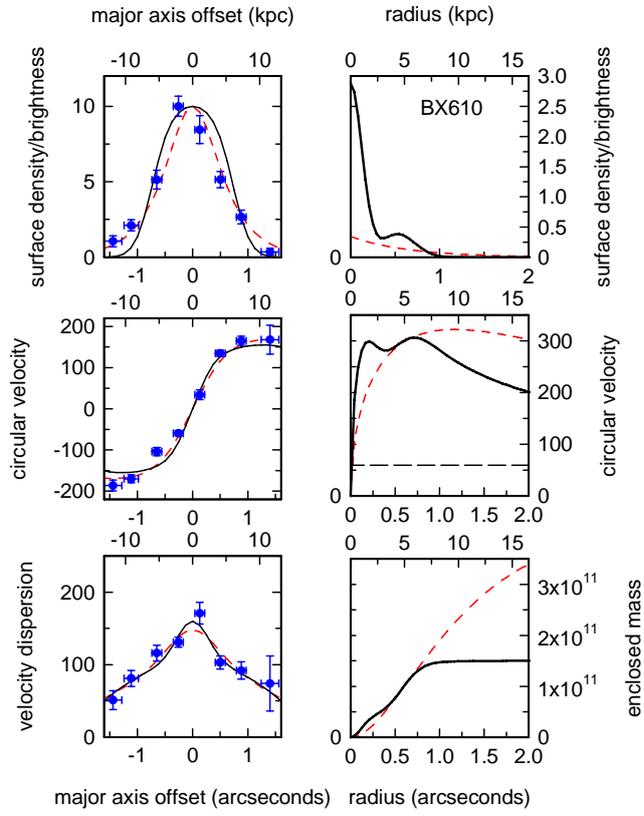

Figure 8. Hα major axis cuts and kinematic modeling of Q2343-BX610 ($z$=2.211), at a resolution of ~0.5" FWHM (see also Förster Schreiber et al. 2006). Symbols and markers are the same as in Figure 3.

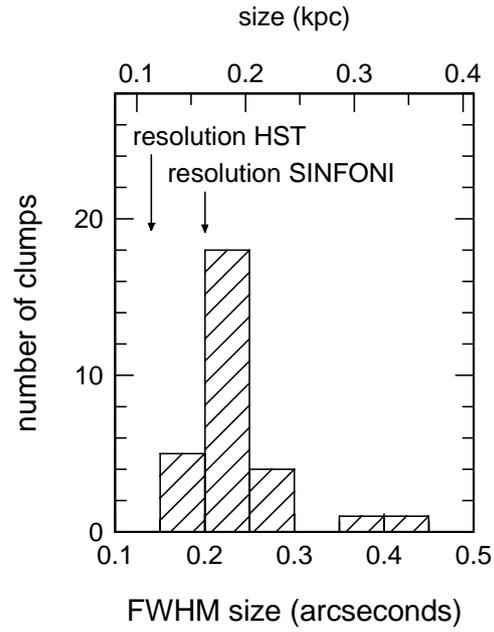

Figure 9. Distribution of observed FWHM sizes of the star-forming clumps in massive star-forming z~2 galaxies. Sizes were determined by Gaussian fits to 29 fairly isolated clumps in the SINFONI data of BzK15504 (Genzel et al. 2006) and BX482 (resolution 0.2"), as well as the NIC2 $H_{160}$ data of BX389, BX610 and MD41 (Förster Schreiber et al. 2008b).

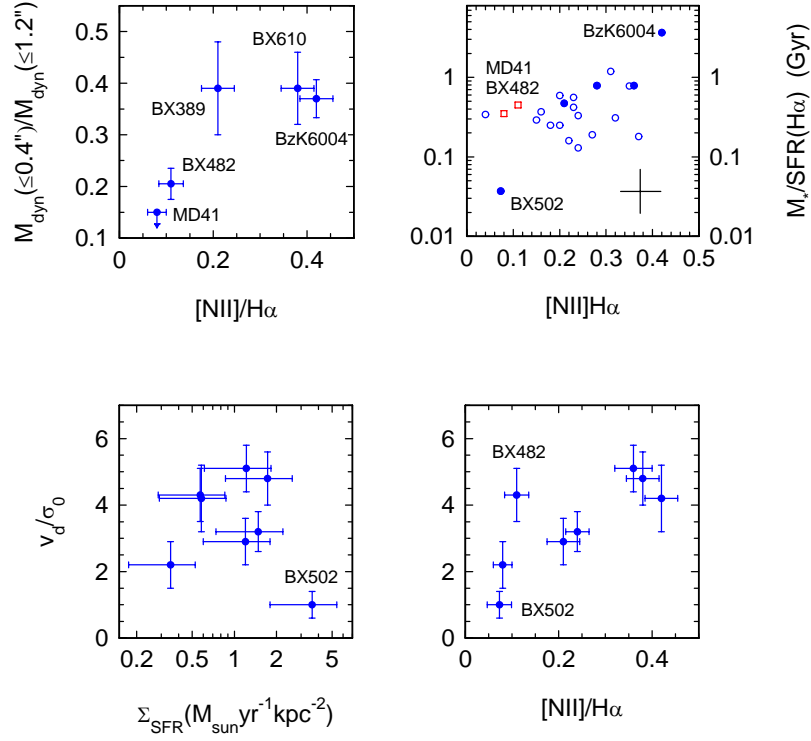

Figure 10. Parameter correlations. Top left: Derived ratio of dynamical masses within 0.4" (~3.3kpc) and 1.2" (~10 kpc) for the five optically/UV-selected star-forming galaxies in Table 2, as a function of the [NII]/Hα flux ratio. Top right: 'star formation age', that is the time to build up the stellar mass at the current (extinction corrected Hα) star formation rate, as a function of [NII]/Hα for the z>2 SINS survey galaxies (open circles), as well as those five galaxies in Table 3 with measured $M_*$/SFR (filled circles and open squares). The large black cross denotes the typical uncertainty. Bottom left: ratio of $v_d/\sigma_0$ as a function of star formation surface density, for the eight sources in Table 3. The compact, dispersion limited galaxy BX502 is marked. Bottom right: $v_d/\sigma_0$ ratio as a function of [NII]/Hα for the eight sources in Table 3.

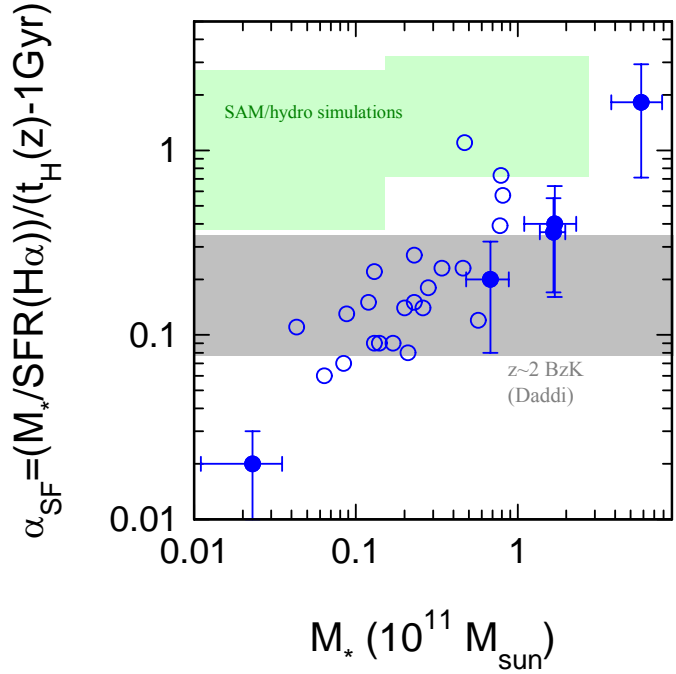

Figure 11. Star formation activity parameter z~2 star-forming galaxies. The grey shaded region are the observed galaxies in the s-BzK survey of Daddi et al. (2007) and green shaded region mark the predictions from current semi-analytic and hydrodynamic galaxy formation simulations as shown in Davé (2007). The best quality SINS galaxies from Förster Schreiber et al. (2008) are the open circles and the filled circles (with 1σ uncertainties) are those five galaxies in Table 3 with a determination of $M_*$. Most of the observed z~2 star-forming galaxies appear to be forming stars at a rate faster than the available Hubble time, in contrast to the predictions from simulations. The trend of $\alpha_{SF}$ with mass for the SINS galaxies is largely a result of the Hα luminosity threshold in the sample (Förster Schreiber et al. 2008).